\documentstyle[aps,prl,multicol,epsfig]{revtex}

\begin{document}

\def\eeq{\end{equation}}
\def\beq{\begin{equation}}
\def\bea{\begin{eqnarray}}
\def\eea{\end{eqnarray}}

\title{ Chaos edges of $z$-logistic maps: Connection between the relaxation and  
sensitivity entropic indices}

\author{Ugur Tirnakli$^{1}$ and 
Constantino Tsallis$^{2,3}$ 
}
\address{
$^1$Department of Physics, Faculty of Science, Ege University, 35100 Izmir, Turkey \\
$^2$Santa Fe Institute, 1399 Hyde Park Road, Santa Fe, New Mexico 87501, USA \\ 
$^3$Centro Brasileiro de Pesquisas F\'\i sicas, 
Rua  Xavier Sigaud 150, 22290-180 Rio de Janeiro, RJ, Brazil \\
}


\maketitle

\begin{abstract}
Chaos thresholds of the $z$-logistic maps $x_{t+1}=1-a|x_t|^z$ $(z>1; t=0,1,2,...)$ 
are numerically analysed at accumulation points of cycles 2, 3 and 5. 
We verify that the nonextensive $q$-generalization of a Pesin-like identity is preserved 
through averaging over the entire phase space. More precisely, we computationally verify 
$\lim_{t \to\infty }\langle S_{q_{sen}^{av}} \rangle(t)/t= \lim_{t \to\infty}\langle \ln_{q_{sen}^{av}} \, 
\xi \rangle(t)/t \equiv \lambda_{q_{sen}^{av}}^{av}$, 
where the entropy 
$S_{q} \equiv \left(1- \sum_i p_i^q\right)/ \left(q-1\right)$ ($S_1=-\sum_ip_i \ln p_i$), 
the sensitivity to the initial conditions 
$\xi \equiv \lim_{\Delta x(0) \to \, 0} \Delta x(t)/\Delta x(0)$, and  
$\ln_q x \equiv \left(x^{1-q}-1\right)/ \left(1-q\right) $ ($\ln_1 x=\ln x$).
The entropic index $q_{sen}^{av}<1$, and the coefficient $\lambda_{q_{sen}^{av}}^{av}>0$  
depend on both $z$ and the cycle. We also study the relaxation that occurs if we start with an ensemble of 
initial conditions homogeneously occupying the entire phase space. The associated Lebesgue measure 
asymptotically decreases as $1/t^{1/(q_{rel}-1)}$ ($q_{rel}>1$). 
These results led to (i) the first illustration of the connection (conjectured by one of us) between 
sensitivity and relaxation entropic indices, namely $q_{rel}-1 \simeq A (1-q_{sen}^{av})^\alpha$, where  
the positive numbers $(A,\alpha)$ depend on the cycle; (ii) an unexpected and new scaling, namely 
$q_{sen}^{av}(cycle\; n)=2.5 \, q_{sen}^{av}(cycle\, 2)+ \epsilon $ ($\epsilon=-0.03$ for $n=3$, 
and $\epsilon = 0.03$ for $n=5$).

\noindent
{\it PACS Number(s): 05.20.-y, 05.45.-a, 05.45.Ac}
\end{abstract}


$ $

\begin{multicols}{2}

Boltzmann-Gibbs (BG) entropy and corresponding statistical mechanics generically require 
{\it strong chaos} for their applicability and (notorious) usefulness. This type of requirement 
was first used in 1872 by Boltzmann himself \cite{Boltzmann1872}. Indeed, his 
``molecular chaos hypothesis" allowed him to arrive to the celebrated distribution of energies 
at thermal equilibrium, now known as {\it Boltzmann weight}. 
Today, we know that this requirement essentially amounts, for classical nonlinear dynamical 
systems, to having at least one {\it positive} Lyapunov exponent. 
In the case of many-body Hamiltonian systems, such a condition is satisfied when the 
interactions are short ranged. Such systems typically exhibit three basic {\it exponential} 
functions \cite{Tsallis2004}, namely (i) the sensitivity to the initial conditions diverges 
exponentially with time, (ii) physical quantities exponentially relax with time to their value 
at the stationary state (thermal equilibrium), and (iii) at thermal equilibrium, 
the probability of a given microstate exponentially decays with the energy of the microstate. 
These three exponentials of different, though related, nature can be summarized in the following 
differential equation:

\begin{equation}
\frac{dy}{dx}=a_1\,y \;\;\;(y(0)=1)\,,
\end{equation}
whose solution is $y=e^{a_1x}$ (the subindex $1$ will become transparent soon). 
Let us make explicit the point. The {\it first} physical interpretation concerns the sensitivity 
to the initial conditions of say a one-dimensional case and is defined as

\beq
\label{sens}
\xi (t) \equiv \lim_{\Delta x(0)\rightarrow 0} \frac{\Delta x(t)}{\Delta x(0)},
\eeq 
where $\Delta x(t)$ is the distance, in phase space, between two copies at time $t$.  
If the system has a positive Lyapunov exponent $\lambda_1$, then $\xi$ diverges as $\xi= e^{\lambda_1 t}$. 
In other words, in this case we have $(x,y,a_1) \equiv (t,\xi,\lambda_1)$. 
The {\it second} physical interpretation concerns the relaxation of some (characteristic) 
physical quantity ${\cal O}(t)$ to its value ${\cal O}(\infty)$ at thermal equilibrium. 
With the definition $\Omega \equiv \frac{{\cal O}(t)-{\cal O}(\infty )}{{\cal O}(0)-{\cal O}(\infty )}$, 
we typically have $\Omega = e^{-t/\tau}$, where $\tau$ is the relaxation time. In other words, 
in this case we have $(x,y,a_1)\rightarrow (t,\Omega ,-1/\tau )$. 
This relaxation occurs precisely because of the sensitivity to initial conditions, 
which guarantees strong chaos. It was apparently Krylov the first to realize \cite{Krylov1944}, 
over half a century ago, this deep connection. The {\it third} physical interpretation is given by 
$p_i=e^{-\beta E_i}/Z$ (with $Z\equiv \sum^W_{j=1}e^{-\beta E_j}$), where $E_i$ is the 
eigenvalue of the $i$-th quantum state of the Hamiltonian (with its associated boundary conditions), 
$p_i$ is the probability of occurrence of the $i$-th state when the system is in equilibrium 
with a thermostat whose temperature is $T\equiv 1/k\beta$ (Gibbs' {\it canonical ensemble}). 
In other words, in this case we have $(x,y,a_1)\rightarrow (E_i,Zp_i,-\beta)$.

A substantially different situation occurs when the maximal Lyapunov exponent vanishes. 
In this case the typical differential equation becomes

\begin{equation}
\frac{dy}{dx}=a_q\,y^q \;\;\;(y(0)=1; q \in {\cal R})\,,
\end{equation}
whose solution is $y=e_q^{a_q \,x}$, the {\it $q$-exponential} function being defined as follows: 
$e_q^x \equiv [1+(1-q)x]^{1/(1-q)}$ if the quantity between brackets is nonnegative, 
and zero otherwise ($e_1^x=e^x$). The sensitivity to the initial conditions is given in this 
case by \cite{TPZ,BaldovinRobledo} $\xi=e_{q_{sen}}^{\lambda_{q_{sen}}\,t}$ ({\it sen} stands for 
{\it sensitivity}). In other words, we have $(x,y,q,a_q) \equiv (t, \xi, q_{sen},  \lambda_{q_{sen}})$. 
The relaxation is typically expected \cite{MouraTirnakliLyra} to be characterized by  
$\Omega = e_{q_{rel}}^{-t/\tau_{q_{rel}}}$ ({\it rel} stands for {\it relaxation}). 
In other words, in this case we have $(x,y,q,a_q) \equiv (t, \Omega, q_{rel}, -1/ \tau_{q_{rel}})$. 
For the longstanding metastable states \cite{LatoraRapisardaTsallis2001} that precede thermal 
equilibrium for long-range interacting Hamiltonians, it is expected \cite{TsallisGell-Mann2004} 
$p_i=e_{q_{stat}}^{-\beta_{q_{stat}} E_i}/Z_{q_{stat}}$  
(with $Z_{q_{stat}} \equiv \sum^W_{j=1}e_{q_{stat}}^{-\beta_{q_{stat}} E_j}$) ({\it stat} 
stands for {\it stationary}). In other words, in this case we have 
$(x,y,q,a_q) \equiv (E_i, Z_{q_{stat}} p_i, q_{stat}, -\beta_{q_{stat}})$. 
For systems that are at, or close to, the edge of chaos we typically have 
$q_{sen} \le 1$, $q_{rel} \ge 1$, and $q_{stat} \ge 1$. For the BG case, where there is one or 
more {\it positive} Lyapunov exponents, we recover  the confluence $q_{sen} =q_{rel} =q_{stat} =1$.

One expects the entire $q$-triplet to be either measurable or calculable for Hamiltonian systems. 
And indeed it has recently been measured in the solar wind \cite{BurlagaVinas-Figueroa}. 
However, the generic relation among these three $q$ indices is still unknown. 
For dissipative systems such as say the $z$-logistic map, no $q_{stat}$ exists. Therefore, the 
problem reduces to only two $q$ indices, namely $q_{sen}$ and $q_{rel}$. Their generic relation 
also is unknown. In the present paper, we provide the first (numerical) evidence of such a connection.

Before entering into the details of the present calculation, let us briefly review the connection 
with the entropy $S_q$, basis of a current generalization of BG statistical mechanics referred to as 
{\it nonextensive statistical mechanics} \cite{Tsallis}. This entropy is defined as follows:

\beq
S_q \equiv \frac{1- \sum_{i=1}^W p_i^q}{q-1} = \sum_{i=1}^W p_i \ln_q (1/p_i)      
\eeq
where the {\it $q$-logarithm} function, inverse of the $q$-exponential, is defined as 
$\ln_q x \equiv \frac{x^{1-q}-1}{1-q} \;(\ln_1 x=\ln x)$ and 
$S_1=S_{BG} \equiv -\sum_{i=1}^W p_i \ln p_i$.

If we partition the phase space of a one-dimensional map (at its edge of chaos) into $W$ small 
intervals, randomly place $N$ initial conditions into one of those windows, and then run the 
dynamics for each of those $N$ points, we get, as time $t$ evolves, an occupancy characterized by 
$\{ N_i(t)\}$ ($\sum_{i=1}^W N_i(t)=N$). 
With $p_i(t) \equiv N_i(t)/N$ we can calculate $S_q(t)$ for any value of $q$. From this, we can 
calculate the {\it entropy production per unit time} \cite{hilborn}, defined as follows:

\beq
K_q \equiv \lim_{t\rightarrow\infty} \lim_{W\rightarrow\infty}
\lim_{N\rightarrow\infty} \frac{S_q(t)}{t}
\eeq
It has been proved \cite{BaldovinRobledo2004} that only $K_{q_{sen}}$ is {\it finite} 
($K_q =0$ for $q>q_{sen}$ and $K_q$ diverges for $q<q_{sen}$). Furthermore, if we consider the 
upper bound of $K_{q_{sen}}$ with regard to the choice of the little window within which we put 
the $N$ initial conditions, we obtain the Pesin-like identity $K_{q_{sen}}=\lambda_{q_{sen}}$.  
Several aspects of this problem have already been verified for various one-dimensional 
unimodal 
maps \cite{lyra,z1z2,TirnakliAnanosTsallis,BorgesTsallisAnanosOliveira}. 
It was recently studied the influence of averaging \cite{AnanosTsallis}. It was verified that, 
while the $q$-generalized Pesin-like identity is preserved, the value of $q_{sen}$ is changed into 
$q_{sen}^{av}$ ({\it av} stands for {\it average}).  
The main goal of the present paper is to exhibit that a simple relation exists between  
$q_{sen}^{av}$ and $q_{rel}$ by making use of the $z$-logistic map family defined as  
 
\beq
x_{t+1} = 1 - a |x_t|^z \;\;\; 
\eeq
where ($z>1; \,0<a \le 2;\, |x_t| \le 1; \, t=0,1,2,...$).

\end{multicols}

\begin{table}
\caption{\label{tab:Table}$z$-logistic map family for cycles 2, 3 and 5.}
\begin{tabular}{||c|c|c|c|c|c|c||} \hline
$z$ & cycle & $a_c$ &  $q_{sen}^{av}$ & $q_{rel}$ & $\lambda_{q_{sen}^{av}}^{av}$ & $K_{q_{sen}^{av}}^{av}$\\ \hline
$1.75$ & $2$ & $1.355060...$ & $0.37\pm 0.01$ & $2.25\pm 0.02$ & $0.26\pm 0.01$ & $0.26\pm 0.02$ \\ \hline
$1.75$ & $3$ & $1.747303...$ & $0.92\pm 0.01$ & $2.25\pm 0.02$ & $0.48\pm 0.01$ & $0.47\pm 0.02$ \\ \hline
$1.75$ & $5$ & $1.607497...$ & $0.96\pm 0.01$ & $2.25\pm 0.02$ & $0.42\pm 0.01$ & $0.40\pm 0.02$ \\ \hline
$2$    & $2$ & $1.401155...$ & $0.36\pm 0.01$ & $2.41\pm 0.02$ & $0.27\pm 0.01$ & $0.27\pm 0.02$ \\ \hline
$2$    & $3$ & $1.779818...$ & $0.88\pm 0.01$ & $2.41\pm 0.02$ & $0.49\pm 0.01$ & $0.48\pm 0.02$ \\ \hline
$2$    & $5$ & $1.631019...$ & $0.93\pm 0.01$ & $2.41\pm 0.02$ & $0.42\pm 0.01$ & $0.40\pm 0.02$ \\ \hline
$2.5$  & $2$ & $1.470550...$ & $0.34\pm 0.01$ & $2.70\pm 0.02$ & $0.28\pm 0.01$ & $0.28\pm 0.02$ \\ \hline
$2.5$  & $3$ & $1.828863...$ & $0.82\pm 0.01$ & $2.70\pm 0.02$ & $0.48\pm 0.01$ & $0.47\pm 0.01$ \\ \hline
$2.5$  & $5$ & $1.669543...$ & $0.88\pm 0.01$ & $2.70\pm 0.02$ & $0.38\pm 0.01$ & $0.37\pm 0.01$ \\ \hline
$3$    & $2$ & $1.521878...$ & $0.32\pm 0.01$ & $2.94\pm 0.02$ & $0.29\pm 0.02$ & $0.29\pm 0.03$ \\ \hline
$3$    & $3$ & $1.862996...$ & $0.78\pm 0.01$ & $2.94\pm 0.02$ & $0.44\pm 0.01$ & $0.44\pm 0.01$ \\ \hline
$3$    & $5$ & $1.699440...$ & $0.84\pm 0.01$ & $2.94\pm 0.02$ & $0.34\pm 0.01$ & $0.35\pm 0.01$ \\ \hline
$5$    & $2$ & $1.645533...$ & $0.28\pm 0.01$ & $3.53\pm 0.03$ & $0.30\pm 0.02$ & $0.30\pm 0.03$ \\ \hline
$5$    & $3$ & $1.931072...$ & $0.68\pm 0.01$ & $3.53\pm 0.03$ & $0.36\pm 0.01$ & $0.37\pm 0.01$ \\ \hline
$5$    & $5$ & $1.773088...$ & $0.73\pm 0.01$ & $3.53\pm 0.03$ & $0.27\pm 0.01$ & $0.25\pm 0.02$ \\ \hline
\end{tabular}
\end{table}

\begin{multicols}{2}

\begin{figure}
\includegraphics[width=0.45\textwidth,keepaspectratio,clip=]{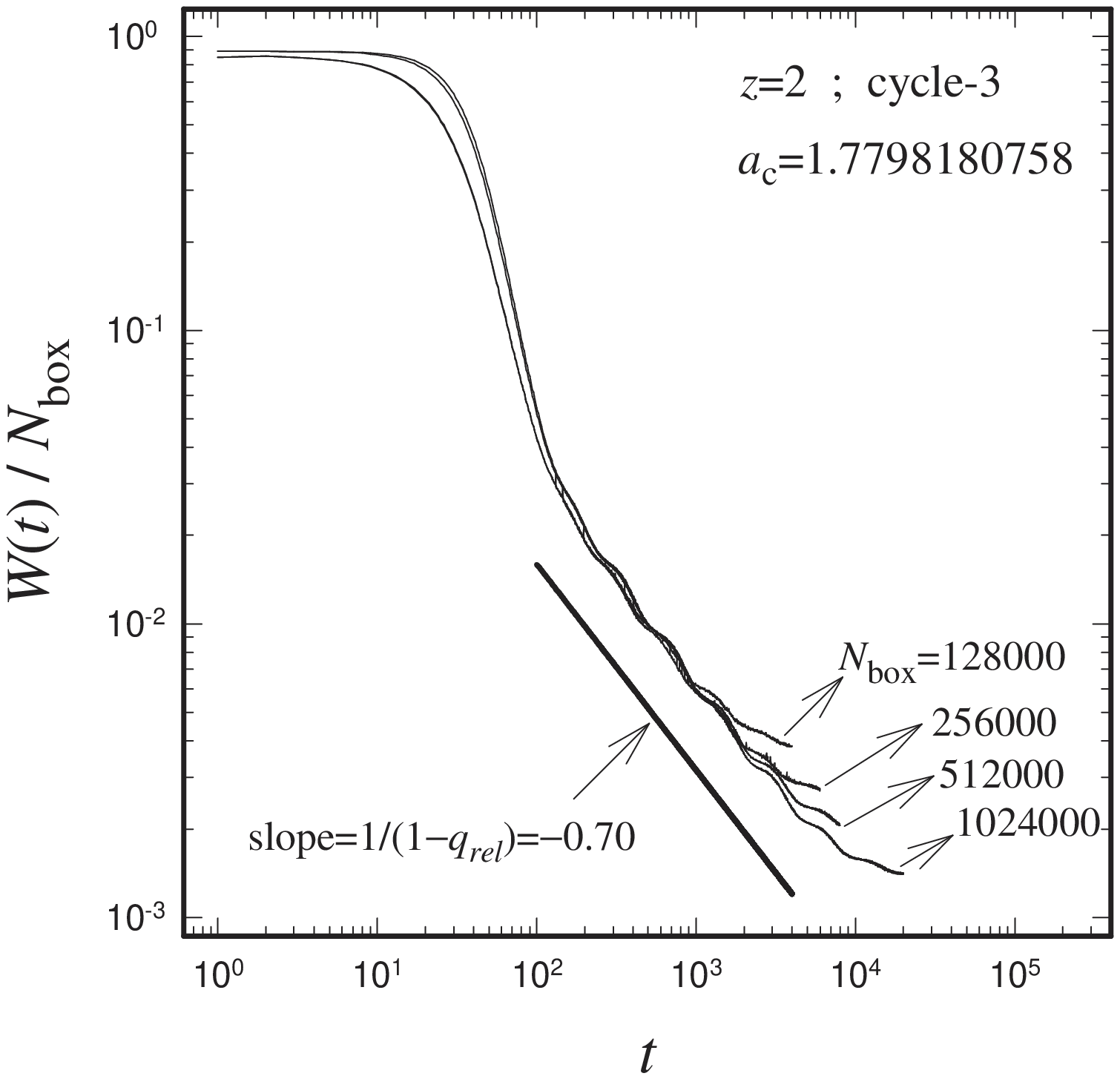}
\caption{\label{fig:Fig1} 
The volume occupied by the ensemble as a function of discrete time. After a transient period, 
which is the same for all $N_{box}$ values, the power-law behavior is evident. For each case, 
the evolution of a set of $10 N_{box}$ identical copies of the system is followed.}
\end{figure}

In our simulations, we firstly check whether $q_{rel}$ values of cyles 3 and 5 are different from 
those of cycle 2 obtained in \cite{MouraTirnakliLyra}. To accomplish this task, we analyse the 
rate of convergence to the critical attractor when an ensemble of initial conditions is uniformly 
distributed over the entire phase space (the phase space is partioned $N_{box}$ cells of equal size) 
and we found that, for all cycles that we studied, the volume $W(t)$ occupied by the ensemble 
exhibits a power-law decay with the same exponent value for fixed $z$. As an example, the case 
of cycle 3 for $z=2$ is given in Fig.~1. The same kind of behavior is obtained also for other 
$z$ values and cycles, which yields us to conclude that $q_{rel}$ does not depend on the cycle 
(see also the Table). 
Then, we concentrate on the ensemble averages of the sensitivity function $\xi(t)$ by considering 
two very close points (throughout this work we take $\Delta x(0)=10^{-12}$) and calculating its 
value from Eq.~(\ref{sens}). This procedure has to be repeated many times with different $x$ values 
randomly chosen in the available phase space and finally an average is taken over all $\ln_q\xi(t)$ 
values. For cycle 3 and cycle 5, we obtain the behavior of $\langle\ln_q\xi\rangle(t)$ as a function 
of $t$, for various $z$ values, from where one can deduce $q_{sen}^{av}$ by identifying the {\it linear} 
time dependence as it is seen in Fig.~2. We verify that $q_{sen}^{av}$ and $\lambda_{q_{sen}^{av}}^{av}$ 
values do depend on both $z$ and the cycle, whereas $q_{rel}$ is independent of the cycles. 
Finally, to investigate the entropy production for cycles 3 and 5, we employ the procedure used 
so far in \cite{TirnakliAnanosTsallis} for cycle 2 of the $z$-logistic 
map family. It is numerically verified that, as seen for a representative case in Fig.~3, for each 
$z$ value and for each cycle, the linear entropy production occurs for a special $q$ value which 
coincides with the one obtained from the sensitivity function. In addition to this, we obtained 
$K_{q_{sen}^{av}}^{av}= \lambda_{q_{sen}^{av}}^{av}$, which clearly broadens the validity region of 
the standard Pesin theorem \cite{pesin}.

\begin{figure}
\includegraphics[width=0.45\textwidth,keepaspectratio,clip=]{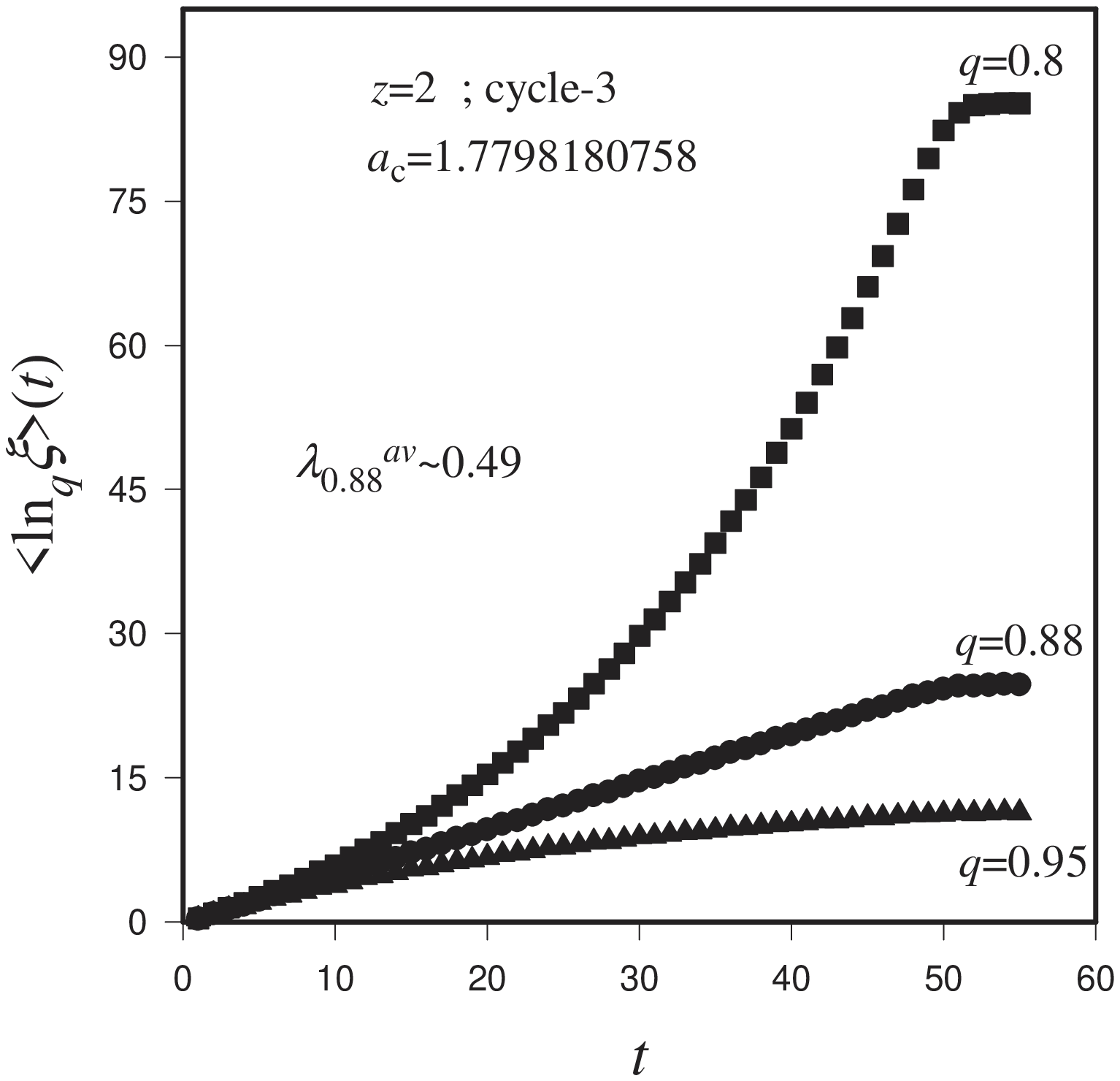}
\caption{\label{fig:Fig2} 
The behavior of $\langle\ln_q\xi\rangle$ as a function of time.}
\end{figure}

\begin{figure}
\includegraphics[width=0.45\textwidth,keepaspectratio,clip=]{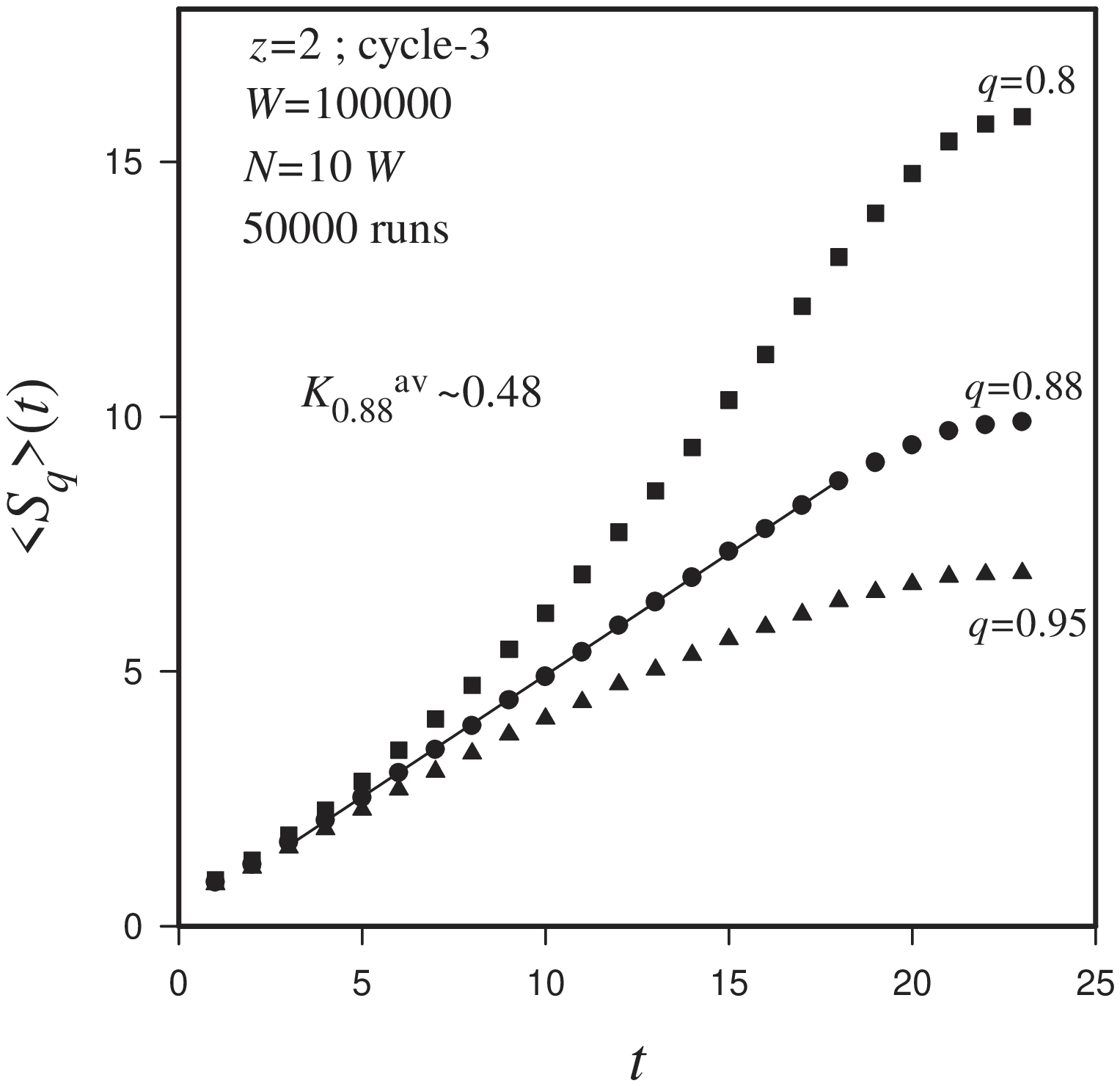}
\caption{\label{fig:Fig3} 
The behavior of $\langle S_q \rangle$ as a function of time.}
\end{figure}

Finally, for all cycles that we studied, we numerically verified that a simple scaling relation 
exists between $q_{sen}^{av}$ and $q_{rel}$ (see Fig.~4): 

\beq
q_{rel}(cycle\; n) -1 \simeq A \left[1-q_{sen}^{av}(cycle\; n)\right]^{\alpha}
\eeq
where $n=2,3,5$ and the values of $A$ and $\alpha$ are given in the caption of Fig.~4; both numbers 
depend on the cycle. For example, $\alpha = 5.1$ for cycle 2, and quickly approaches zero when the 
cycle increases; $A$ also decreases when the cycle increases. This kind of relation between these 
two classes of $q$ index is seen for the first time in a model system. It is clearly consistent with 
the confluence occurring for BG systems. This is to say, when there is at least one positive Lyapunov 
exponent, we obtain $q_{rel}=q_{sen}^{av}=1$.

We also notice (see Fig.~5) a new and unexpected scaling behavior, namely 
\begin{equation}
q_{sen}^{av}(cycle\; n)=2.5 \, q_{sen}^{av}(cycle\, 2)+ \epsilon \,,
\end{equation}
with $\epsilon=-0.03$ for $n=3$, and $\epsilon = 0.03$ for $n=5$. 

Summarizing, we have discussed a paradigmatic family of one-dimensional dissipative maps, and have shown that 
its (averaged) sensitivity to the initial conditions and its relaxation in phase space follow a simple 
path, which is consistent with current nonextensive statistical mechanical concepts, and which 
considerably extends the validity of Pesin-like identities. The sensitivity to the initial conditions 
is characterized by $q_{sen}^{av}<1$, which monotonically approaches unity with increasing cycle size 
(at least for the specific cycles that we have studied here), and decreases with $z$. It is further 
characterized by $\lambda_{q_{sen}^{av}}^{av}$, which exhibits a maximum both as a function of the 
cycle size and of $z$. The relaxation is characterized by $q_{rel}>1$, which monotonically increases 
with $z$ and does not depend on the cycle. This study has enabled to exhibit two interesting relations, 
namely Eqs.~(7) and (8). This path is expected to appreciably enlighten, among others, the case of 
long-range-interacting Hamiltonian systems, where the situation is even more complex since a {\it third} 
entropic index, $q_{stat}$, is expected, which would characterize the energy distribution at metastable 
states. Analytic analysis of the scalings presented here are certainly most welcome.

\begin{figure}
\includegraphics[width=0.45\textwidth,keepaspectratio,clip=]{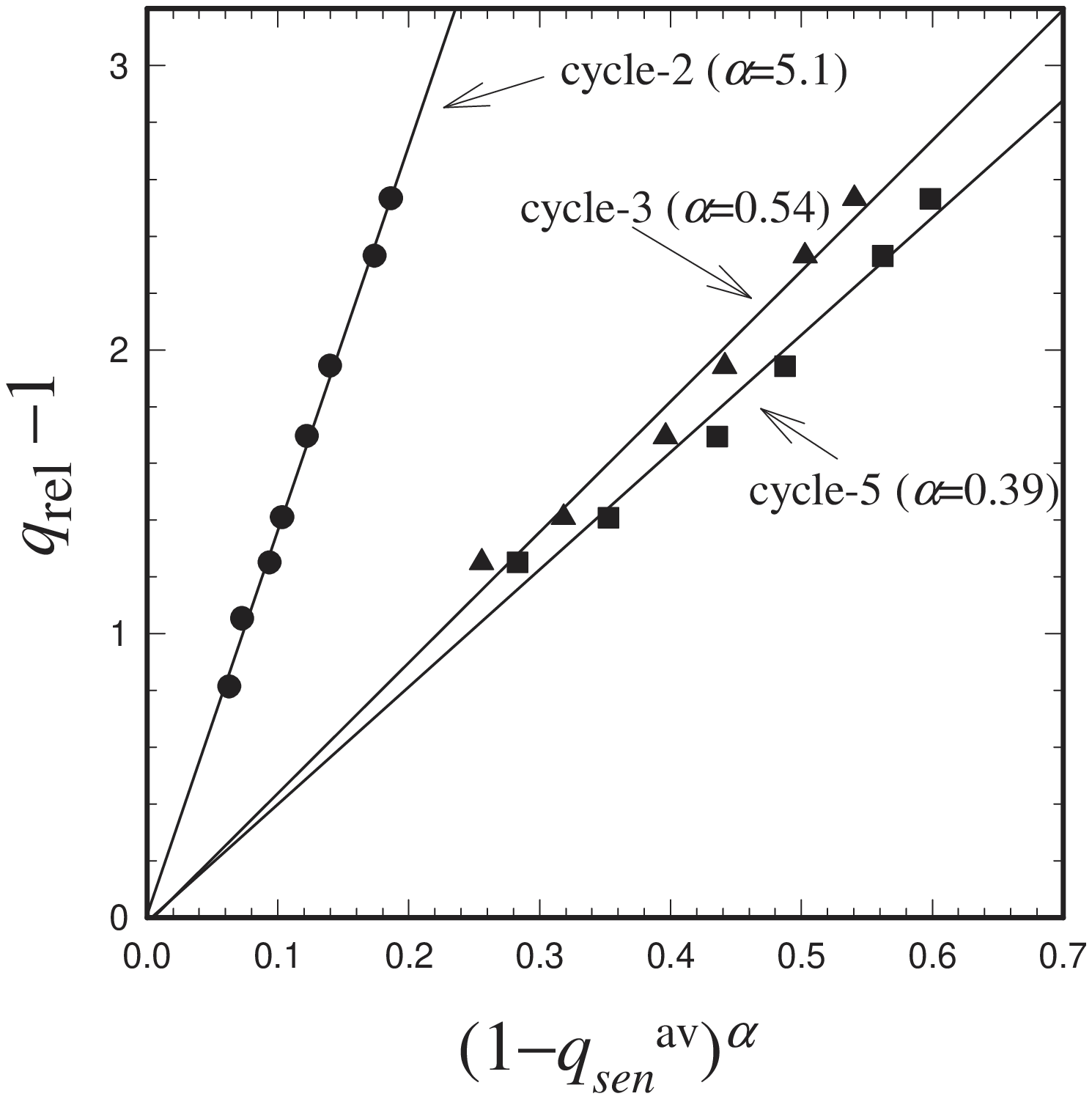}
\caption{\label{fig:Fig4} 
Straight lines: $q_{rel}(cycle \, 2) -1 =  13.5 \, [1  
-   \\
q_{sen}^{av}(cycle \,2)]^{5.1}$, 
$q_{rel}(cycle \, 3) -1 = 
4.6 \, [1-q_{sen}^{av}(cycle \,3)]^{0.54}$, and 
$q_{rel}(cycle \, 5) -1 = 4.1 \, [1-q_{sen}^{av}(cycle \,5)]^{0.39}$. }
\end{figure}

\begin{figure}
\includegraphics[width=0.45\textwidth,keepaspectratio,clip=]{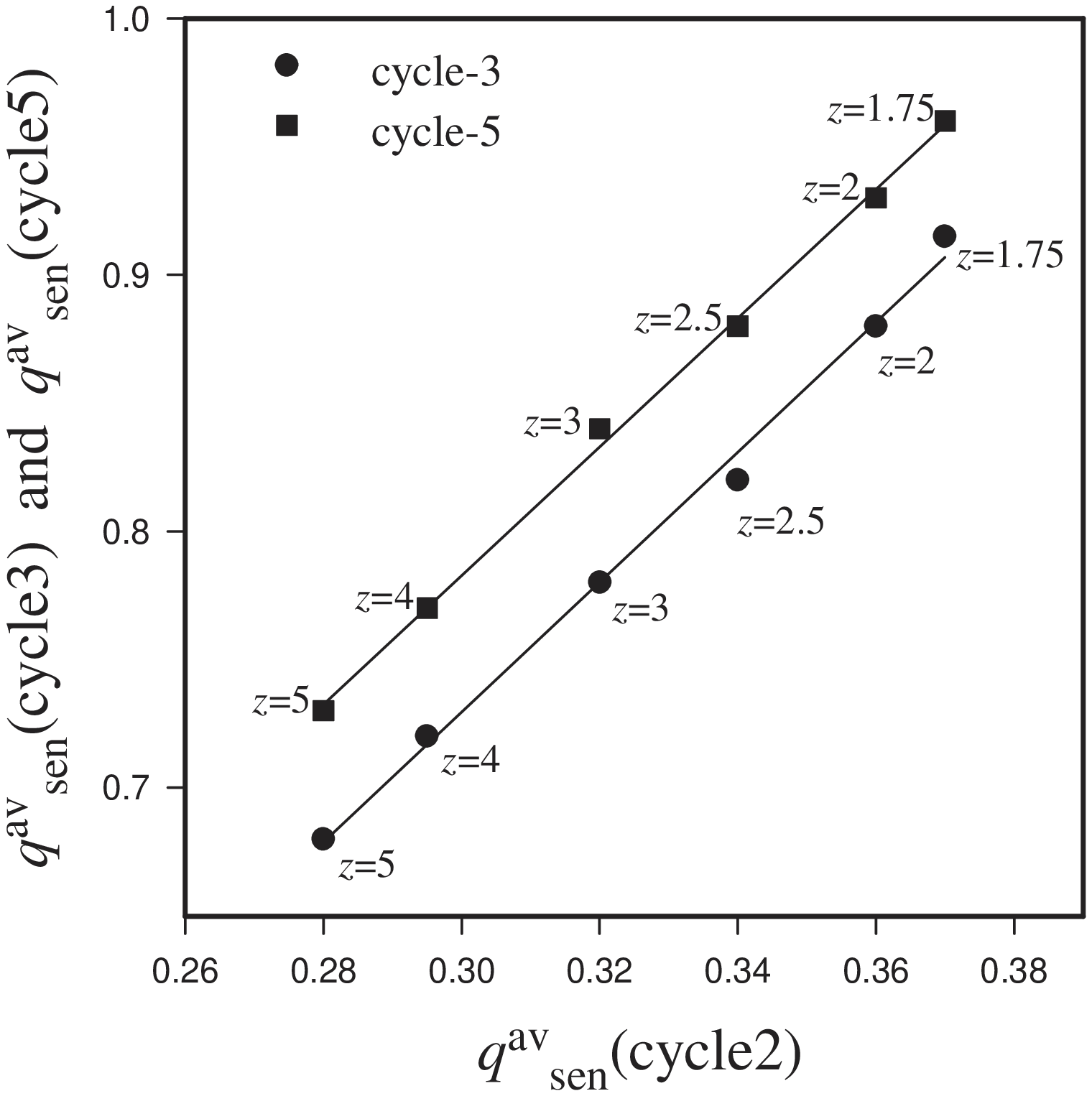}
\caption{\label{fig:Fig5} Straight lines: $q_{sen}^{av}(cycle\, 3)= 
2.5 \,q_{sen}^{av}(cycle \,2)  \\
-0.03$, and 
$q_{sen}^{av}(cycle \,5)=
2.5 \, q_{sen}^{av}(cycle \,2) 
+0.03$, which suggests 
$q_{sen}^{av}(cycle \,5)-q_{sen}^{av}(cycle \,3) \simeq 0.06$. }
\end{figure}

This work is partially supported by TUBITAK (Turkish agency), Pronex, CNPq and Faperj 
(Brazilian agencies). 

\end{multicols}

\end{document}